\documentclass[aps,prb,twocolumn,superscriptaddress]{revtex4-2}
\usepackage{bbm}
\usepackage{amsmath}
\usepackage{amssymb}
\usepackage{graphicx}
\usepackage{mathrsfs}
\usepackage{amsfonts}
\usepackage{amsthm}
\usepackage{color}
\usepackage{txfonts}
\usepackage{lipsum}
\usepackage[normalem]{ulem}
\usepackage{gensymb}
\usepackage{booktabs}
\usepackage[colorlinks=true,citecolor=blue,linkcolor=blue,urlcolor=blue,anchorcolor=blue]{hyperref}

\begin{document}

\title{Twisted magnon frequency combs in ferromagnetic nanorings}

\author{Xuejuan Liu}
\affiliation{College of Applied Sciences, Shenzhen University, Shenzhen 518060, China}
\affiliation{Shenzhen Key Laboratory of Ultraintense Laser and Advanced Material Technology, Center for Intense Laser Application Technology, and College of Engineering Physics, Shenzhen Technology University, Shenzhen 518118, China}

\author{Xingen Zheng}
\affiliation{The Center for Advanced Quantum Studies and School of Physics and Astronomy, Beijing Normal University, Beijing 100875, China}
\affiliation{Key Laboratory of Multiscale Spin Physics, Beijing Normal University, Beijing 100875, China}

\author{Zhengyi Li}
\affiliation{School of Physics and State Key Laboratory of Electronic Thin Films and Integrated Devices, University of Electronic Science and Technology of China, Chengdu 610054, China}

\author{Zhizhi Zhang}
\affiliation{School of Mechanical and Electrical Engineering, Chengdu University of Technology, Chengdu 610059, China}

\author{Xiaoguang Li}
\affiliation{Shenzhen Key Laboratory of Ultraintense Laser and Advanced Material Technology, Center for Intense Laser Application Technology, and College of Engineering Physics, Shenzhen Technology University, Shenzhen 518118, China}

\author{Hai-Peng Sun}
\email[Corresponding author: ]{hpsun@sztu.edu.cn}
\affiliation{Shenzhen Key Laboratory of Ultraintense Laser and Advanced Material Technology, Center for Intense Laser Application Technology, and College of Engineering Physics, Shenzhen Technology University, Shenzhen 518118, China}

\author{Hui Li}
\email[Corresponding author: ]{lihui@sztu.edu.cn}
\affiliation{Shenzhen Key Laboratory of Ultraintense Laser and Advanced Material Technology, Center for Intense Laser Application Technology, and College of Engineering Physics, Shenzhen Technology University, Shenzhen 518118, China}

\author{Cangtao Zhou}
\email[Corresponding author: ]{zcangtao@sztu.edu.cn}
\affiliation{College of Applied Sciences, Shenzhen University, Shenzhen 518060, China}
\affiliation{Shenzhen Key Laboratory of Ultraintense Laser and Advanced Material Technology, Center for Intense Laser Application Technology, and College of Engineering Physics, Shenzhen Technology University, Shenzhen 518118, China}

\begin{abstract}
We systematically investigate the emergence of twisted magnon frequency combs (tMFCs) and their higher-order modes arising from strong nonlinear coupling between vortex-core gyration and azimuthal spin-wave modes in ferromagnetic nanorings. The frequency-comb spacing is set by the gyrotropic frequency, which can be tuned by both the size of the central hole and external magnetic fields. Remarkably, for the larger hole diameter (50 nm), an additional magnon mode emerges, leading to additional tMFC families. We also demonstrate that the selection rules still hold for different sizes of the central hole. In addition, the external in-plane magnetic field provides an effective means to control the tMFC, while the response depends on the nanostructure geometry. The nanodisk shows an approximately symmetric response under field reversal, whereas the response of nanorings depends on the size of the central hole. For a small hole diameter (5 nm), the low-field response becomes asymmetric, and the frequency-comb spacing increases with field magnitude over the higher-field branches. A larger hole diameter (50 nm) raises the gyrotropic frequency, yielding a sparser sideband structure near the drive frequency. Our results show that ferromagnetic nanorings support geometrically and magnetically tunable tMFCs. 
\end{abstract}

\maketitle

\section{INTRODUCTION}
Magnetic vortices are fundamental topological spin structures at the mesoscopic scale, characterized by in-plane magnetization curling around a perpendicularly magnetized core \cite{Bader2006,Yu2021,Shinjo2000, Wachowiak2002,LiNANO2021}. Their configuration is defined by the core polarity ($P=\pm1$) and chirality ($C=\pm1$) \cite{Guslienko2008a,Jaafar2010}. The dynamic properties of magnetic vortices are governed by three classes of excitations: a sub-GHz gyrotropic mode, higher-frequency azimuthal spin-wave modes, and higher-frequency radial spin-wave modes  \cite{Back2004,Buess2005,Philippe2026,Novosad2005, Ivanov2005, Taurel2016}. The cylindrical symmetry of the system gives rise to the quantization of the azimuthal spin-wave modes (or twisted magnons), which are consequently labeled by the radial index $n=0,1,2,\dots$ and the azimuthal index $l=0,\pm1,\pm2,\dots$ \cite{Guslienko2002,Salama2025}. The clockwise rotation is defined as negative, while counterclockwise rotation is considered positive. More importantly, the twisted magnon carries orbital angular momentum (OAM), associated with a helical phase factor $e^{i l\phi}$ in its wave function, where $\phi$ is the azimuthal angle and the integer $ l $ denotes the OAM quantum number~\cite{Li2022,Jia2019,Jia2019NC}.

The interaction between twisted spin waves and magnetic vortices manifests distinctly across different driving regimes. In the weakly driven regime,  the vortex core (VC) mediates the splitting of azimuthal spin-wave modes~\cite{Park2005,Guslienko2008,Verba2021}. Besides, spin-wave excitation can also induce VC reversal~\cite{Kammerer2011,Yoo2015}. In the strongly driven regime, vortex dynamics become
highly nonlinear, enabling the excitation of whispering-gallery magnons \cite{Schultheiss2019} and generation of twisted magnon frequency combs (tMFCs) \cite{Wang2022,Gao2023}. These combs arise from three-magnon confluence and splitting processes that couple the gyrating VC to azimuthal spin-wave modes \cite{Aristov2016,Zhang2018,Jenkins2021}. Specifically, the steady-state gyration of the VC at a fixed radius periodically modulates the local effective field, which in turn imposes a time-periodic modulation on the spin waves. This periodic renormalization of the magnon manifold produces Floquet sidebands, which manifest as the characteristic tMFC spectrum \cite{Heins2026,Seeger2026}. 

Extending this concept to nanorings opens up new possibilities beyond the nanodisk geometry. An important question is how the central hole modifies tMFC formation, the associated selection rules, and the response to an in-plane magnetic field. Recently, Heins et al. found that active control of magnon Floquet states and MFC generation can be achieved by tuning the nucleation and gyrotropic motion of the vortex core using the excitation power and an in-plane magnetic field~\cite{Heins2025}. Furthermore, controlled symmetry breaking provides an effective means of tailoring magnetic-vortex dynamics \cite{Iurchuk2024,Saavedra2025}.

In this work, we systematically investigate the generation and evolution of tMFCs and their higher-order modes in nanorings by means of micromagnetic simulations. Our results reveal that the nanoring geometry preserves the selection rule, while the frequency-comb spacing can be tuned via the size of the central hole and in-plane magnetic fields. Notably, when the hole diameter is increased from 5 nm to 50 nm, additional tMFC branches emerge due to an additional magnon mode, and the number of combs within the same frequency range increases. In addition, applying external in-plane magnetic fields results in field-direction-dependent
spectral asymmetry and tunes the frequency-comb spacing. These findings establish ferromagnetic nanorings as a versatile platform for nonlinear magnonics with potential applications in magnonic sensing and precision metrology.

The paper is organized as follows: Sec. \ref{Model and Micromagnetic Simulations} presents the model and methods used in this work. Section \ref{Results and Discussion} shows the spectra and selection rules of the tMFCs in ferromagnetic nanorings, with a particular focus on how an external in-plane magnetic field controls the frequency-comb spacing. Section \ref{Conclusion} summarizes the key results of this work.

\section{Model and Micromagnetic Simulations} \label{Model and Micromagnetic Simulations}

We perform micromagnetic simulations using the open-source software Mumax$^{3}$  \cite{Vansteenkiste2014,Vanderveken2021}, which numerically solves the  Landau-Lifshitz-Gilbert (LLG) equation \cite{Landau1935,Gilbert2004}:
\begin{equation}\label{Eq1}
\begin{aligned}
\frac{\partial \mathbf{m}}{\partial t}=-\gamma\mathbf{m}\times {\mathbf{H}_{\texttt{eff}}}+\alpha\mathbf{m}\times\frac{\partial\mathbf{m}}{\partial{t}},
\end{aligned}
\end{equation}
where $\mathbf{m}$ is the unit magnetization vector, $\gamma$ is the gyromagnetic ratio, $\mathbf{H}_{\texttt{eff}}$ is the effective field, $\alpha$ is the Gilbert damping constant, and $t$ is time. The effective field is given by $\mathbf{H}_{\texttt{eff}}=-\frac{1}{\mu_{0}M_{s}} \frac{\delta W}{\delta \mathbf{m}}$, with the vacuum permeability $\mu_0$, the saturation magnetization $M_{s}$ and the total magnetic energy $W$. $W$ consists of the exchange energy $E_{\texttt{ex}}$, the anisotropy energy $E_{\texttt{an}}$, the magnetostatic energy $E_{\texttt{m}}$, and the Zeeman energy $ E_{\texttt{Zee}}$.

\begin{figure}[t]
\begin{centering}
\includegraphics[width=0.48\textwidth]{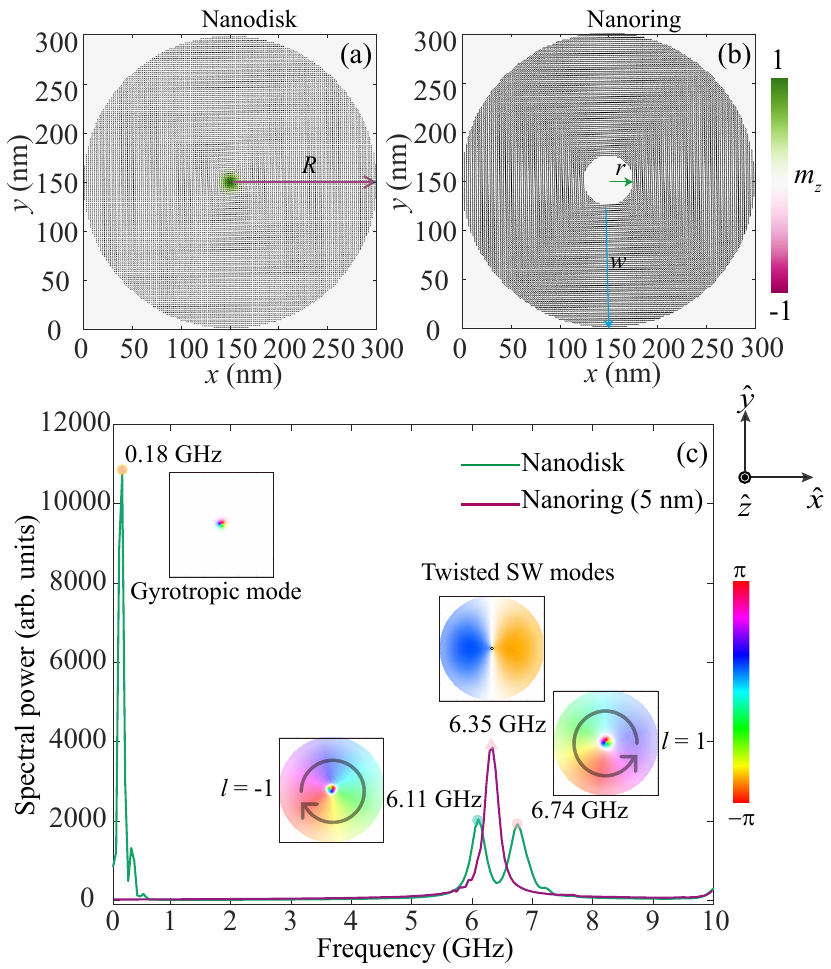}
\par\end{centering}
\caption{(a) Spatial configuration of $m_z$ in a nanodisk with diameter $2R = 300$ nm and thickness $d = 5$ nm. (b)~The corresponding $m_z$ configuration in a nanoring with hole diameter $2r = 50$~nm. (c) Spectral power of the nanodisk (green) and nanoring with $2r=5$ nm (purple) under excitation by an in-plane sinc-shaped magnetic field $\mathbf{B}_\mathrm{sinc}(t)$. The inset displays spatial maps of $\delta m_{z}$ at the resonance frequencies 0.18, 6.11, 6.35 and 6.74 GHz. The yellow dot represents the gyrotropic mode, while the green and pink dots denote the azimuthal spin-wave modes with $l = -1$ and $l = +1$, respectively. The pink triangle indicates the degenerate mode of azimuthal spin waves with $|l| = 1$ in the nanoring. }
\label{Figure1}
\end{figure}

Material parameters for Permalloy (Py) were adopted \cite{Wang2022}: the exchange constant $A_{\texttt{ex}}=1.3\times10^{-11} $ J/m, the saturation magnetization $M_{s}=8.6\times10^{5} $ A/m, and the damping constant $\alpha = 0.01$ unless otherwise stated. The cell size was set to $1 \times 1 \times 5$~nm$^{3}$ with open boundary conditions. The system is initialized in a vortex state. Figure~\ref{Figure1}(a) shows a ferromagnetic nanodisk with outer diameter $2R=300$~nm and thickness $d=5$~nm. Figure~\ref{Figure1}(b) illustrates the nanoring geometry with hole diameter $2r$ and ring width $w = R -r$. We investigate the vortex dynamics in the nanodisk and nanorings. 

To characterize the vortex excitation spectra, we apply a spatially uniform, in-plane sinc-shaped magnetic field:
\begin{equation}\label{EqSinc}
\mathbf B_{\mathrm{sinc}}(t)=B_0
\frac{\sin[2\pi f_c(t-t_0)]}{2\pi f_c(t-t_0)}\,\hat{\mathbf x},
\end{equation}
with the amplitude $B_{0} =5$~mT, time offset $t_{0}=1$~ns, and cutoff frequency $f_{c} = 20$~GHz. The total simulation time is 40 ns, with a time interval of 5 ps. We perform a Fourier transformation (FT) at each cell to extract the eigenmodes from the excitation spectra \cite{Wanghao2007, BuessPRL2005,LiuAIP2020}. The magnetization is decomposed into discrete Fourier components as

\begin{equation}
m_j(t) = \sum_{\omega} m_{j,\omega}(t)
= \sum_{\omega} \text{Re}\left[m_{j,\omega} e^{i\omega t}\right],
\end{equation}
where the superscript $j \in \{x, y, z\}$ labels the Cartesian component of the magnetization, and $\omega = 2\pi f$ with $f$ the frequency.
The corresponding Fourier coefficient is written as:
\begin{equation}
m_{j,\omega} = \left| m_{j,\omega} \right| e^{i\phi_{j,\omega}},
\end{equation}
with $\left| m_{j,\omega}\right|$ denoting the spectral amplitude and $\phi_{j,\omega}$ the spectral phase. Through this site-resolved FT analysis, the two-dimensional spatial distribution of each resonant mode is obtained at its characteristic eigenfrequency.
In the subsequent analysis, we focus primarily on the $m_z$ component to illustrate the spectral characteristics.

\section{Results and Discussion} \label{Results and Discussion}

\subsection{Formation of tMFCs in the small-hole nanoring}

By applying the fast Fourier transform to $\delta m_z$ over all magnetic cells, we obtain the spectra presented in Fig.~\ref{Figure1}(c). For the nanodisk [green line in Fig.~\ref{Figure1}(c)], three prominent peaks are resolved at 0.18~GHz, 6.11~GHz, and 6.74~GHz, corresponding to the gyrotropic mode, the $l=-1$ azimuthal
spin-wave mode, and the $l=+1$ azimuthal spin-wave mode,
respectively. The gyrotropic mode arises from the translational oscillation of the VC around the equilibrium position at the nanodisk center, and a weak second harmonic is also observable at 0.36~GHz. 

By contrast, for the nanoring, we perform calculations for a hole diameter of $2r=5$~nm to efficiently explore the parameter space and establish the underlying physics. In the nanoring [purple line in Fig.~\ref{Figure1}(c)], the gyrotropic mode is fully suppressed, leaving a single spin-wave mode at 6.35~GHz with $|l| = 1$. This suppression confirms that the splitting of azimuthal modes originates from the dipolar interaction between azimuthal spin waves and the VC \cite{Guslienko2008,LiuAIP2020,Hoffmann2007}.

\begin{figure}[b]
\begin{centering}
\includegraphics[width=0.5\textwidth]{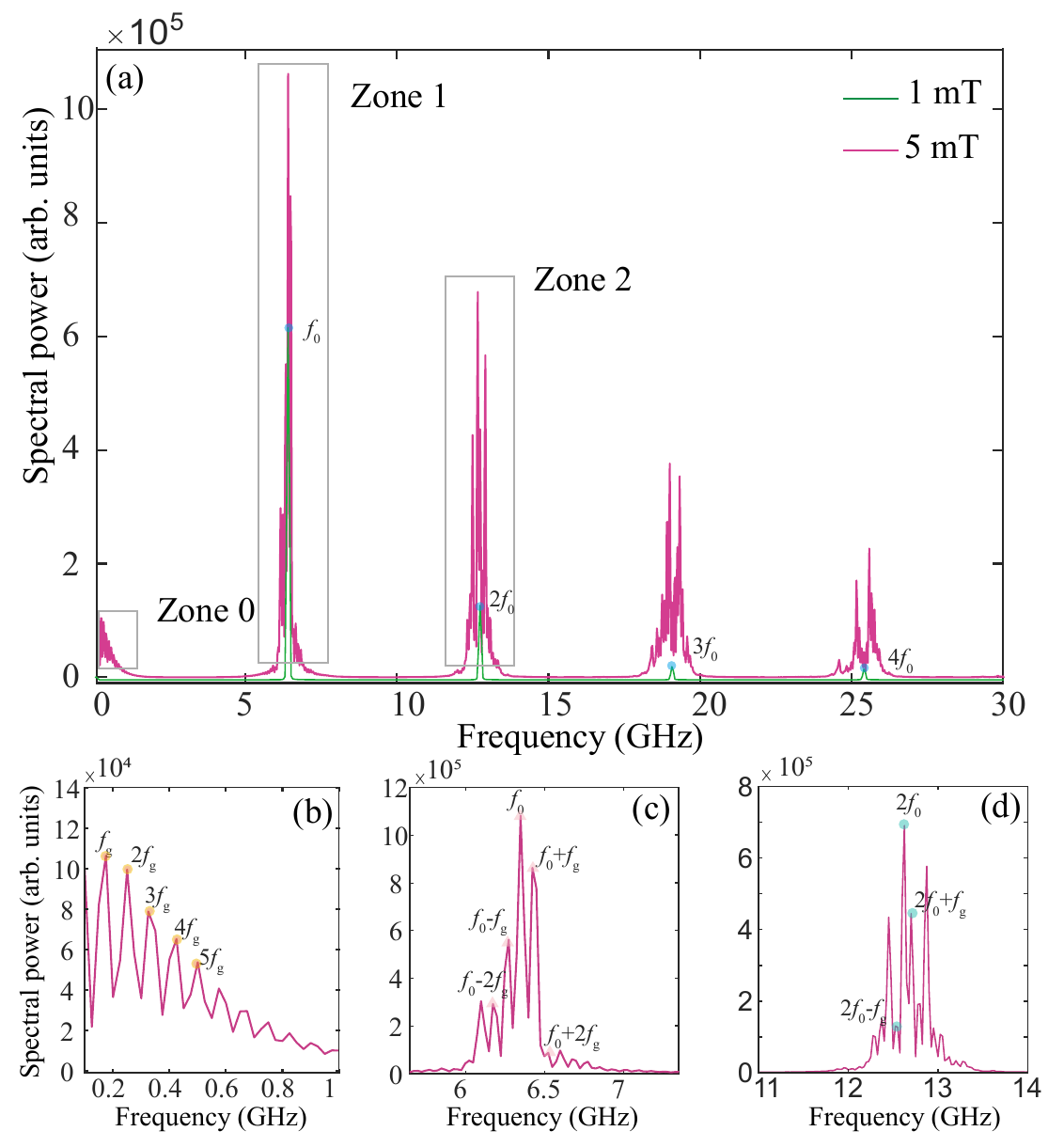}
\par\end{centering}
\caption{(a) FFT spectra of the nanoring with hole diameter $2r = 5$~nm under rotating magnetic fields $\mathbf B_{\mathrm{rf}}$ with amplitudes $B_0 = 1$~mT and $B_0 = 5$~mT at driving frequency $f_0 = 6.35$~GHz. (b)--(d) Magnified views of the three boxed regions in (a), where (b), (c) and (d) correspond to Zones 0, 1 and 2, respectively. }
\label{Figure2}
\end{figure}

Based on the eigenfrequencies of the internal excitation modes in the nanoring, an in-plane rotating magnetic field $\mathbf B_{\mathrm{rf}}$ is applied, where
\begin{equation}\label{EqDrive}
\mathbf B_{\mathrm{rf}}(t)=B_0[\cos(2\pi f_0t)\hat{\mathbf x}
+\sin(2\pi f_0t)\hat{\mathbf y}],
\end{equation}
with $f_0=6.35$ GHz to resonantly drive the $l = +1$ azimuthal spin-wave mode. Figure~\ref{Figure2}(a) shows the spectral power of the nanoring under different field amplitudes. For the amplitude $B_{0}=1$~mT [green line], the spectrum is dominated by the directly driven mode at 6.35 GHz and its higher-order harmonics at 12.7, 19.05, and 25.4 GHz. The low-frequency response is weak and no well-developed sideband sequence is resolved. For $B_{0}=5$~mT [pink line], the spectrum changes substantially.
Beyond the directly driven mode at $f_0$ and its higher-order harmonics, additional spectral peaks appear at $f_g$ and at the sum- and difference-frequency sidebands $f_0 \pm f_g$, constituting a characteristic signature of a tMFC.
The three boxed zones in Fig.~\ref{Figure2}(a) are shown in magnified views in Figs.~\ref{Figure2}(b)--(d). Figure~\ref{Figure2}(b) resolves the emergence of $f_g$ at 0.18 GHz and its harmonics 0.36, 0.54, 0.72 GHz and so on, while Figs.~\ref{Figure2}(c) and~\ref{Figure2}(d) reveal the first and second tMFCs [Zone 1 and Zone 2 in Fig.~\ref{Figure2}(a)], respectively. This spectral structure indicates nonlinear coupling between the VC gyration and the azimuthal spin-wave mode. The re-emergence of the gyrotropic response at this field may be associated with the displacement of the VC from the central hole, which restores a confining potential. Efficient comb generation, in turn, requires the VC to follow a circular trajectory that spatially overlaps the region of maximum azimuthal-mode amplitude, thereby enabling nonlinear energy exchange between the two excitations.

\begin{figure}[t]
\centering
\includegraphics[width=0.5\textwidth]{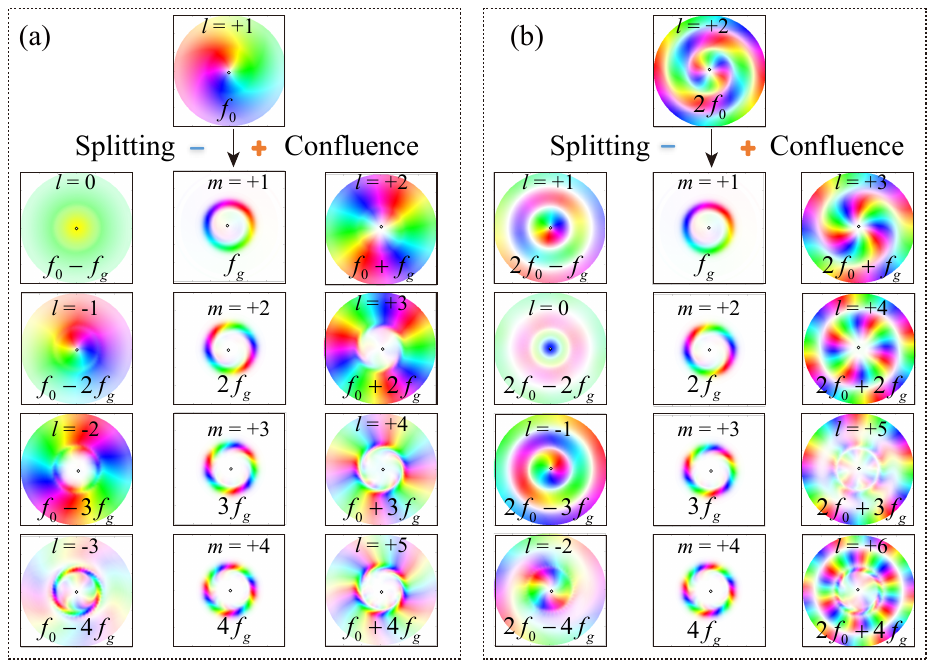}
\caption{Selection rules for the tMFC spectral lines in the nanoring with hole diameter $2r = 5$~nm. (a) Selection rules for the first tMFC shown in Zone 1 in Fig.~\ref{Figure2}(a). (b) Selection rules for the second tMFC shown in Zone 2 in Fig.~\ref{Figure2}(a). Small black circles mark the holes. 
}
\label{Figure3}
\end{figure}

To confirm that the tMFC lines carry distinct OAM quantum numbers, we examine the spatial phase distributions of the gyrotropic mode and the various comb components in Zones 0, 1 and 2 of Fig.~\ref{Figure2} (a).  
Figure~\ref{Figure3} shows the spatial part of the sideband assignment. To avoid confusing a harmonic order with the radial index $n$, we denote the drive-harmonic order by $p$ and the nonnegative
sideband order by $m$ ($m=0,1,2,\ldots$). $m=0$ denotes the common central component.
The plus and minus signs denote the upper- and lower-frequency
branches relative to each comb center, respectively. The illustrated tMFC families satisfy:
\begin{equation}\label{EqComb}
f_{p, m}^{\pm}=p f_0 \pm m f_g,\quad l=p \pm m.
\end{equation}
For $p=1$, the difference-frequency sequence has $l=0,-1,-2,-3$ for $m=1,2,3,4$, while the sum-frequency sequence has $l=2,3,4,5$. For $p=2$, the corresponding difference- and sum-frequency sequences are $l=1,0,-1,-2$ and $l=3,4,5,6$, respectively. These are the labels displayed in Figs.~\ref{Figure3}(a) and \ref{Figure3}(b). 
These relations are consistent with frequency and azimuthal-index-matching conditions for splitting and confluence processes, followed by successive mixing events \cite{Verba2021,Wang2022}. Higher-order sidebands can be generated through successive three-magnon confluence and splitting processes, rather than through a single higher-order scattering event. Moreover, frequency and OAM matching specify the selection rules for the nonlinear scattering processes, but do not by themselves determine the corresponding coupling strength. Because the illustrated modes have different radial profiles, the selection rule for the azimuthal OAM index should not be interpreted as implying conservation of the radial mode index.

\begin{figure}[b]
\centering
\includegraphics[width=0.48\textwidth]{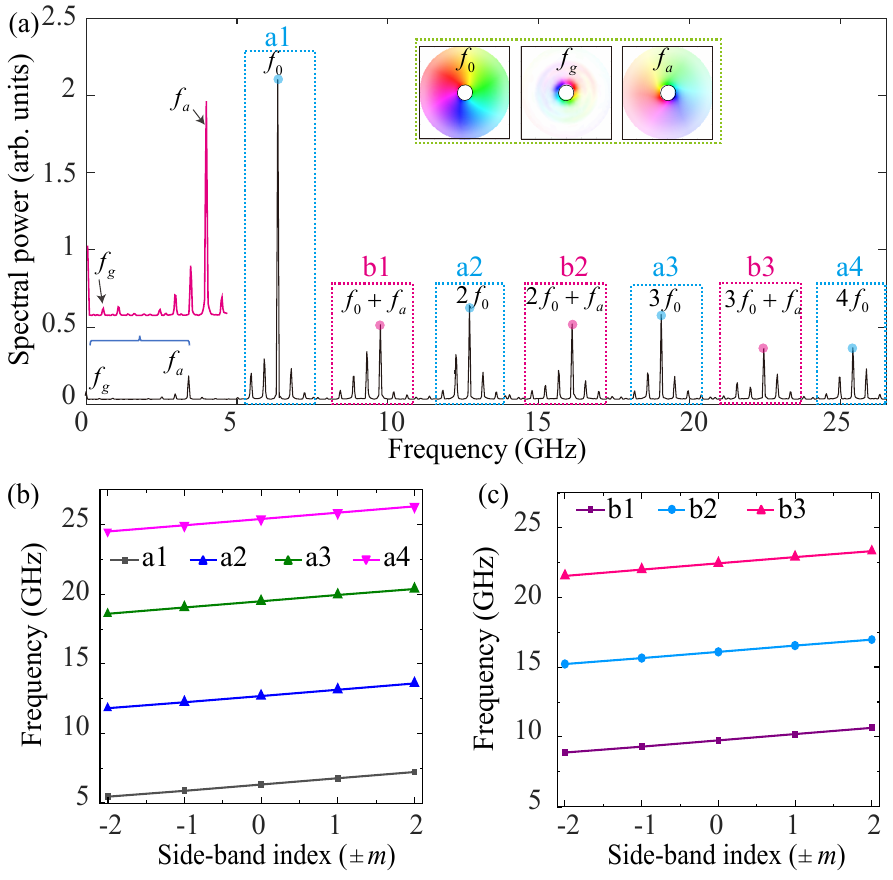}
\caption{(a) FFT spectra of the nanoring with hole diameter $2r = 50$~nm under a rotating magnetic field $\mathbf B_{\mathrm{rf}}$ with amplitude $B_0 = 5$~mT at frequency $f_0 = 6.35$~GHz. The blue and pink boxes indicate the conventional and additional tMFC families, respectively. (b) and (c) show the comb frequencies of the a1-a4 modes (b) and b1-b3 modes (c) versus the side-band order $m$, respectively. $m=0$ denotes the common central component. The plus and minus signs denote the upper- and lower-frequency
branches relative to each comb center, respectively.
}
\label{Figure4}
\end{figure}

\subsection{Additional tMFCs in the large-hole nanoring}

We now investigate the effect of hole size on the response, focusing on a larger central-hole diameter of 
$2r=50$ nm.
Figure~\ref{Figure4}(a) shows the FFT spectrum under a rotating magnetic field $\mathbf B_{\mathrm{rf}}$ with $B_0 = 5$~mT and $f_0 = 6.35$~GHz. The spectrum is dominated by the directly driven mode at $f_0$ and higher-order harmonics $2f_0$, $3f_0$, $4f_0$, located at 12.70, 19.05, and 25.40~GHz, respectively. Compared to the nanoring with a smaller hole ($2r = 5$~nm), the spectrum for  $2r = 50$~nm exhibits a richer comb structure as highlighted by the pink dashed box.

Notably, the gyrotropic frequency increases from $f_g = 0.18$~GHz to $f_g = 0.45$~GHz as the hole diameter increases from $2r = 5$~nm to $2r = 50$~nm. This behavior can be attributed to the hole-induced modification of the VC confinement potential. For a small hole, the VC remains close to the nanodisk center, where the effective potential is relatively shallow and the restoring force is weak, resulting in a low $f_g$. As the hole becomes larger, the VC is increasingly confined to the annular region near the hole edge, where the effective potential is steeper. The resulting enhancement of the restoring force leads to a higher gyrotropic frequency~\cite{Thiele1973,Lee2007}.
In addition, the larger hole introduces a stronger structural perturbation, giving rise to an additional resonance at $f_a = 3.4$~GHz. This additional magnon mode interacts with the fundamental mode $f_0$ and its harmonics ($2f_0$, $3f_0$, $\dots$) via nonlinear processes, generating a series of components ($f_0 + f_a$, $2f_0 + f_a$, $3f_0 + f_a$, and so on). 
The two sets of tMFCs can be represented as:
\begin{align}
f^{(a,\pm)}_{p, m}&=p f_0 \pm m f_g, &&p=1,2,3,4,\label{EqA}\\
f^{(b,\pm)}_{p, m}&=p f_0+f_a \pm m f_g, &&p=1,2,3.\label{EqB}
\end{align}

For the additional tMFCs, the frequency centers are approximately 9.75, 16.10, and 22.45 GHz as shown in Fig.~\ref{Figure4}(a) [marked by pink dots for b1-b3 modes]. 
This additional mode introduces extra spectral lines between the drive harmonics without changing the characteristic frequency-comb spacing $f_g$ within each comb family. 
The resulting spectrum comprises offset comb families, each with spacing $f_g$, and is generally not uniformly spaced as a whole.
This picture explains why the large-hole nanoring can exhibit a richer broadband spectrum while retaining a relatively sparse sideband structure near the drive frequency.

As summarized in Figs.~\ref{Figure4}(b) and (c),
the comb frequencies follow the sideband relations
in Eqs.~\eqref{EqA} and \eqref{EqB},
with adjacent orders within each branch separated by $f_g$.
These frequency relations are consistent with nonlinear
mixing involving $f_0$, $f_g$, and the additional mode $f_a$.

\begin{figure}[t]
\centering
\includegraphics[width=0.48\textwidth]{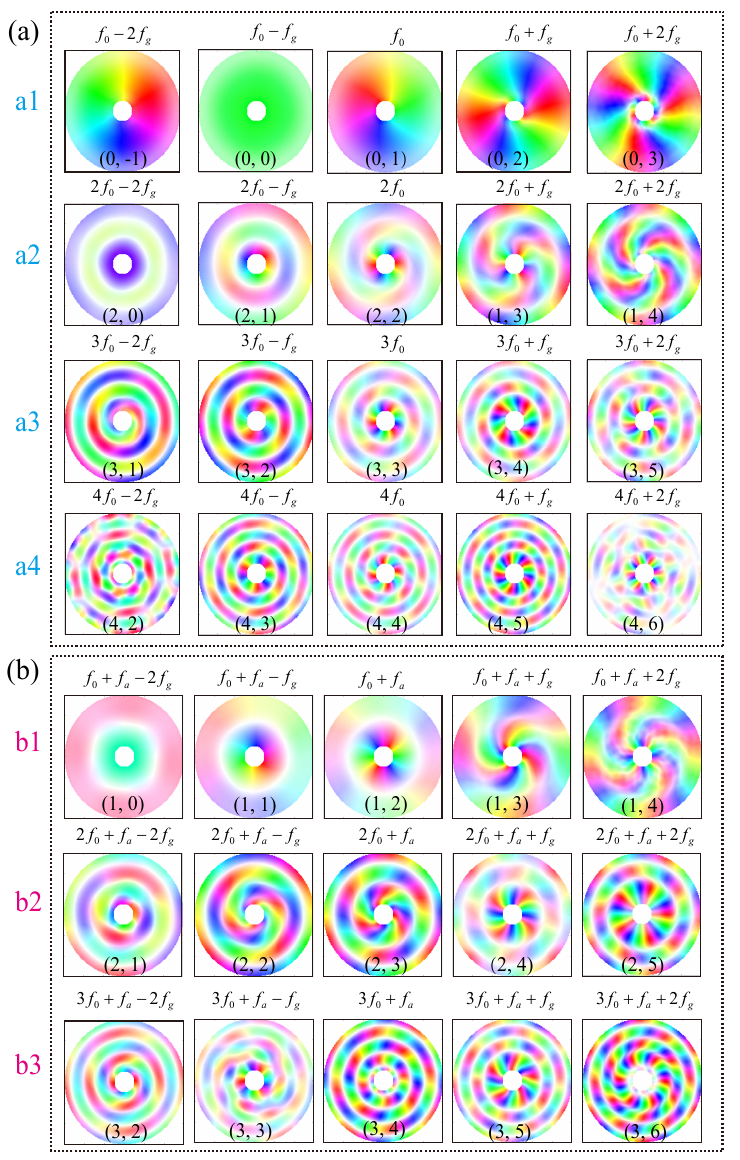}
\caption{Real-space magnetization dynamics of tMFC spectral components for the conventional (a1-a4) and additional (b1-b3) comb families. (a) Sideband modes generated by nonlinear coupling of the drive $f_0$ and its harmonics with the gyrotropic mode $f_g$. 
(b) Sideband modes arising from nonlinear coupling among $f_0$, $f_g$, and the additional mode $f_a$. 
The color scale encodes the azimuthal angle of the in-plane magnetization. $(n,l)$ denote the radial and azimuthal mode indices.}
\label{Figure5}
\end{figure}

The spatial profiles of the tMFC components identified  in Fig.~\ref{Figure4}(a) are presented in Fig.~\ref{Figure5}. Each comb line corresponds to a distinct spin-wave mode with well-defined azimuthal symmetry. Figures~\ref{Figure5}(a) and~\ref{Figure5}(b) illustrate the  spatial profiles of the tMFCs corresponding to the a1-a4 (blue boxes) and b1-b3 (pink boxes) comb families in Fig.~\ref{Figure4}(a), respectively. The spatial profiles shown in Fig.~\ref{Figure5}(a) originate from nonlinear processes that couple $f_0$, $f_g$, and the associated harmonic modes. In contrast, the profiles in Fig.~\ref{Figure5}(b) arise from nonlinear interactions involving the additional mode $f_a$, together with the $f_0$- and $f_g$-family modes, giving rise to a qualitatively distinct spectral structure. Higher drive harmonics $p f_0$ introduce concentric radial nodes, yielding finer spatial structures at elevated frequencies. The additional $f_a$ mode hybridizes with these patterns to produce concentric ring-like modulations, providing a direct spatial signature of the additional nonlinear coupling channel inferred from the spectra in Fig.~\ref{Figure4}(a).

For the eccentric nanorings considered here, displacing the central hole suppresses the tMFCs. This suppression may be associated with changes in the azimuthal mode structure and nonlinear coupling caused by the geometric asymmetry~\cite{Saavedra2025,Streib2021}.

\subsection{Magnetic-field control of tMFCs in nanostructures}
\begin{figure*}[ht]
\begin{centering}
\includegraphics[width=0.8\textwidth]{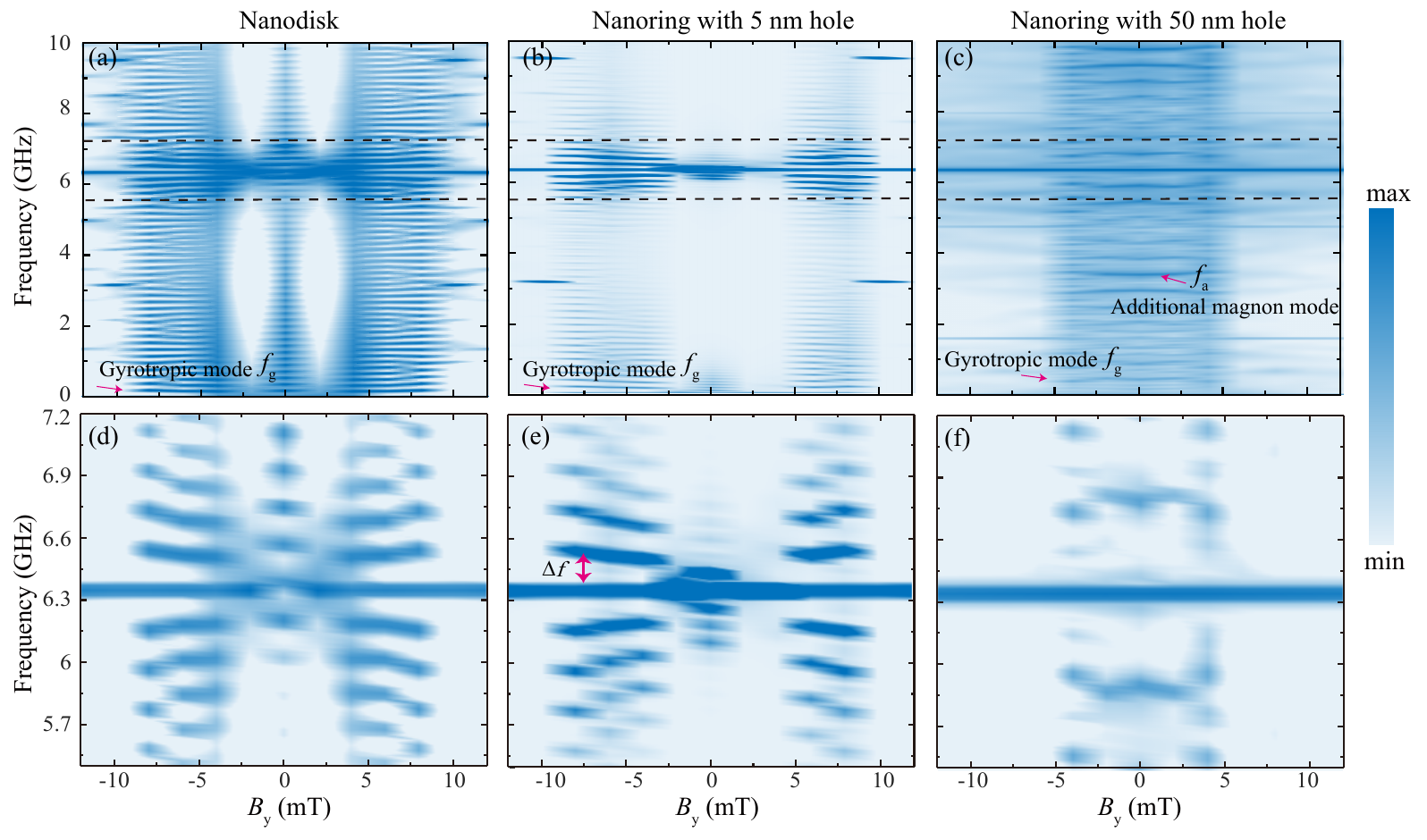}
\end{centering}
\caption{Spectra simulated as a function of a static in-plane magnetic field $B_y$ for different geometries: (a) nanodisk, (b) nanoring with 5 nm hole, (c) nanoring with 50 nm hole. 
The excitation frequency and amplitude were fixed at $f_{0} = 6.35$ GHz and $B_{0}=5$ mT, respectively. (d--f) Magnified views of the dashed boxes (5.5--7.2~GHz) in (a--c), respectively. The sidebands are separated by $\Delta f$. }
\label{Figure6}
\end{figure*}

To investigate the interplay between the tMFC and an external magnetic field, we apply a static in-plane magnetic field $B_y$ to the system. For each field value, the relaxed state is initially determined by energy minimization, following which the magnetization dynamics are acquired under a fixed rotating magnetic field $\mathbf B_{\mathrm{rf}}$ with amplitude $B_0 = 5$~mT and frequency $f_0 = 6.35$~GHz. The external field $B_y$ is swept from $-12$ to $12$~mT in steps of $2$~mT. 
Figure~\ref{Figure6} compares the field-dependent spectra of the nanodisk and nanorings with hole diameters $2r=5$~nm and $2r=50$~nm.

For the nanodisk [Figs.~\ref{Figure6}(a) and \ref{Figure6}(d)], the spectra remain approximately symmetric under reversal of the external magnetic field ($B_y \rightarrow -B_y$). 
Sidebands are visible at zero field and within finite-field windows on both sides of $B_y=0$, with weaker intensity in the intermediate regions. Along the well-defined branches at approximately $4\lesssim |B_y|\lesssim 8$~mT, the sidebands shift away from the drive frequency as $|B_y|$ increases. The applied field therefore affects both the sideband spacing and spectral intensity. Since the evolution varies across the field range, however, it cannot be described by a single monotonic dependence on $B_y$.

The field dependence of the frequency-comb spacing is also evident in the nanoring with hole diameter $2r=5$~nm [Figs.~\ref{Figure6}(b) and \ref{Figure6}(e)]. Unlike the nanodisk, however, the low-field response is no longer symmetric with respect to $B_y=0$. In the positive-field range $2~\mathrm{mT}\lesssim B_y\lesssim4~\mathrm{mT}$, both the gyrotropic mode and the tMFC sidebands are suppressed, consistent with annihilation of the VC~\cite{Heins2025}. In the corresponding negative-field range, $-4~\mathrm{mT}\lesssim B_y\lesssim-2~\mathrm{mT}$, the VC remains stable and the comb structure persists. This asymmetry indicates a field-direction-dependent stability of the VC and, consequently, a direction-dependent switching of the tMFC in the low-field regime. At larger field magnitudes, approximately $4\lesssim |B_y|\lesssim8$~mT, the sidebands move progressively away from the drive frequency, corresponding to a monotonic increase in the frequency-comb spacing with increasing $|B_y|$ as shown in Fig.~\ref{Figure6}(e).

For the nanoring with the larger central-hole diameter ($2r=50$~nm), the field-dependent response becomes more complex, with noticeably different low-frequency dynamics [Fig.~\ref{Figure6}(c)]. The gyrotropic mode shifts to $f_g=0.45$~GHz and shows a reduced spectral intensity, while an additional mode appears at $f_a=3.4$~GHz. This extra mode introduces another channel that may contribute to the nonlinear dynamics and modify the tMFC spectrum. As shown in Fig.~\ref{Figure6}(f), the resulting tMFC is considerably sparser and exhibits a larger mode spacing than those of the nanodisk and the nanoring with the smaller hole.

\section{Conclusion} \label{Conclusion}
We have demonstrated that ferromagnetic nanorings support tMFCs. The frequency-comb spacing is set by the gyrotropic frequency $f_g$, which can be controlled by both the size of the central hole and external magnetic fields. For a small central hole, sufficiently strong driving restores the gyrotropic motion of the VC and enables nonlinear coupling with the azimuthal spin-wave mode, giving rise to tMFCs that follow the selection rules. For a large central hole, the gyrotropic frequency increases and an additional mode at $f_a=3.4$ GHz emerges, resulting in a richer broadband spectrum rather than a single uniformly spaced comb. The external in-plane magnetic field provides an effective means to tune the tMFC, while the response strongly depends on the nanostructure geometry. The nanodisk exhibits an approximately symmetric field dependence, whereas the nanorings show a clear sensitivity to the size of the central hole. For the nanoring with hole diameter $2r=5$~nm, the low-field response becomes asymmetric due to the field-dependent stability of the VC, while the frequency-comb spacing increases with $|B_y|$ over the higher-field branches. When the hole diameter is increased to $2r=50$~nm, the gyrotropic frequency increases and an additional mode $f_a$ appears, resulting in a sparser tMFC spectrum with larger mode spacing. These results show that nanostructure geometry and external magnetic fields provide effective means to tailor tMFCs, opening new opportunities for magnonic sensing and precision metrology.

\begin{acknowledgments}
We thank Peng Yan, Zhenyu Wang and  Tijjani Abdulrazak for helpful discussions. This project was supported by the National Natural Science Foundation of China (Grant No. 12104322), Guangdong Basic and Applied Basic Research Foundation (Grant No. 2025A1515011895), and the Natural Science Foundation of Top Talent of SZTU (Grant No. GDRC202309). Z. Z. Zhang acknowledges support from the NSFC under Grant No. 12404122 and Sichuan Science and Technology Program under Grant No. 2025ZNSFSC0868.
\end{acknowledgments}

\end{document}